\documentclass[conference]{IEEEtran}
\IEEEoverridecommandlockouts
\usepackage{cite}
\usepackage{amsmath,amssymb,amsfonts}
\usepackage{caption,subcaption}
\usepackage{algorithmic}
\usepackage{graphicx}
\usepackage{textcomp}
\usepackage{xcolor}
\usepackage{booktabs}
\usepackage{array}
\usepackage{graphicx}
\usepackage{tikz}
\usepackage{fancyhdr}

\fancypagestyle{IEEEheader}{
    \fancyhf{} 
    
    \fancyhead[L]{%
        \begin{minipage}{\textwidth}
            \centering
            \footnotesize
            This work has been submitted to IEEE for possible publication. \\
            Copyright may be transferred without notice, after which this version may no longer be accessible.\\[3pt]
            \rule{\textwidth}{0.2pt} 
        \end{minipage}
    }
}

\def\BibTeX{{\rm B\kern-.05em{\sc i\kern-.025em b}\kern-.08em
    T\kern-.1667em\lower.7ex\hbox{E}\kern-.125emX}}
    
\begin{document}

\title{CRSF: Enabling QoS-Aware Beyond-Connectivity Service Sharing in 6G Local Networks\\

}

\author{
\IEEEauthorblockN{
    Pragya Sharma\IEEEauthorrefmark{1},
    Amanda Xiang\IEEEauthorrefmark{2},
    Abbas Kiani\IEEEauthorrefmark{2},
    John Kaippallimalil\IEEEauthorrefmark{2},
    Tony Saboorian\IEEEauthorrefmark{2}, 
    Haining Wang\IEEEauthorrefmark{1}
}
\IEEEauthorblockA{
\IEEEauthorrefmark{1}Department of Electrical and Computer Engineering, Virginia Tech, Alexandria, VA, USA\\
\IEEEauthorrefmark{2}Wireless Research and Standards, Futurewei Technologies Inc., Addison, TX, USA
}
}

\maketitle
\thispagestyle{IEEEheader}

\begin{abstract}
Sixth-generation (6G) networks are envisioned to support interconnected local subnetworks that can share specialized, beyond-connectivity services. However, a standardized architecture for discovering and selecting these services across network boundaries has not existed yet. To address this gap, this paper introduces the Central Repository and Selection Function (CRSF), a novel network function for the 6G core that facilitates efficient inter-subnetwork service discovery and selection. We formulate the selection process as a QoS-aware optimization problem designed to balance service quality metrics with user-defined priorities. We evaluate our system model through simulations for a sensing service scenario and observe a consistently higher aggregate Quality of Service (QoS) compared to the baseline selection strategy. The proposed CRSF provides a foundational and extensible mechanism for building standardized, collaborative, and service-centric interconnected networks essential for the 6G era.

\end{abstract}


\section{Introduction}

As global connectivity advances beyond the fifth generation (5G), the telecommunications industry and research communities are actively exploring the use cases and demands for the sixth generation (6G) networks. This global effort is being formalized by standardization bodies like the 3\textsuperscript{rd} Generation Partnership Project (3GPP), which are already laying the groundwork to finalize the first 6G rollout by 2030~\cite{Ericsson_IMT2030}.

\begin{figure}[t!]
    \centering
    \includegraphics[width=\linewidth, trim={1cm 0.6cm 1cm 1cm}, clip]{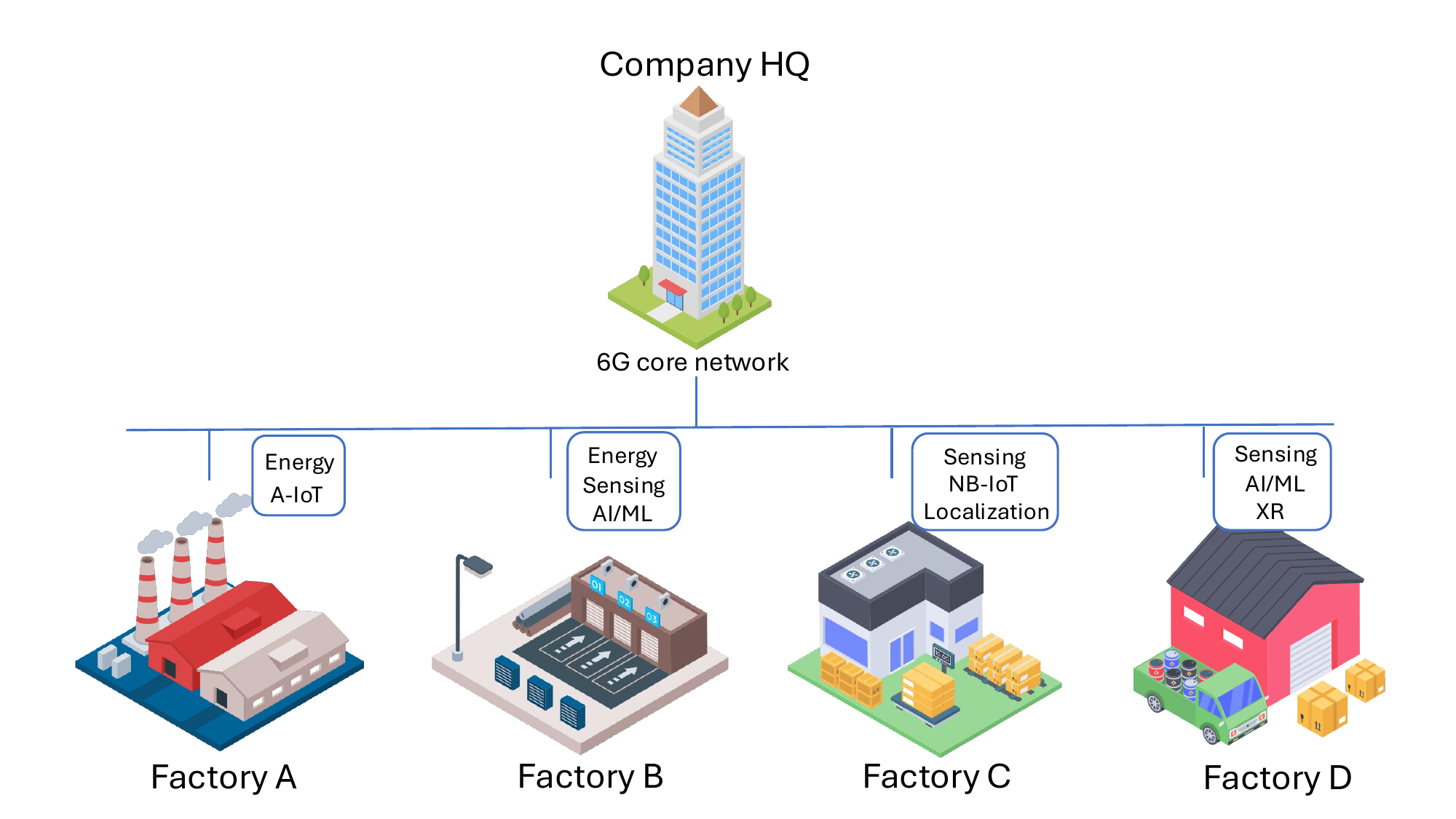}
    \caption{Example of a local 6G network composed of interconnected factories alongside company headquarters.}
    \label{fig:vertical_usecase}
\end{figure}

A key driver for 6G standardization is the concept of local area networks composed of interconnected subnetworks~\cite{6Gusecases}. These subnetworks represent geographically distributed, independently administered networks that collaborate to deliver specialized services essential for industry verticals and smart city ecosystems.
3GPP’s recent study on 6G use cases and service requirements~\cite{3gppTR22870} also highlights this vision, particularly in: (i) Sec. 5.9.2 -- 6G Local Area Networks  and (ii) Sec. 11.9 -- 6G Localized Networks for Verticals. 
An example of such a scenario is shown in Figure~\ref{fig:vertical_usecase} which depicts a multi-site factory network interconnected with its company headquarters, with different services offered across each campus. 
Consequently, the standardization of interfaces, protocols, and architectures that enable seamless interoperability across these interconnected local networks has emerged as a key research direction towards realizing the 6G vision.

Beyond traditional connectivity, some subnetworks may offer advanced services such as Integrated Sensing and Communications (ISAC), artificial intelligence and machine learning (AI/ML) analytics, and high-precision localization, which may not be natively available in other subnetworks. Therefore, a fundamental requirement for the 6G architecture is to enable the efficient sharing of these \textit{beyond-connectivity} services across subnetwork boundaries, allowing the entire system to leverage its collective potential.

This vision of enabling beyond-connectivity service sharing across distributed subnetworks in 6G introduces a design gap not addressed by existing 5G standards. The closest 5G counterpart, Non-Public Networks (NPNs)~\cite{5gnpn}, are primarily designed for isolated deployments, offering tailored services within a self-contained environment but lacking mechanisms for cross-domain collaboration.
While 6G seeks to retain the autonomy and customization benefits of NPNs, it must incorporate mechanisms to support inter-subnetwork service exchange for interoperability and ubiquitous connectivity.

To address the challenge of beyond-connectivity service sharing, the first step is to establish an efficient service discovery and selection mechanism. In this work, we propose an enhanced 6G Service-Based Architecture (SBA) with a novel network function, the \textbf{Central Repository and Selection Function (CRSF)}, which allows consumers to discover the subnetwork offering the desired service and selects the most suitable service provider for consumers. Functionally, the CRSF evolves the concept of the 5G Network Repository Function (NRF) for a group of subnetworks. While the NRF maintains a service repository for a single subnetwork, the CRSF expands this role to operate across multiple domains by collecting and exposing information on beyond-connectivity service functions, providing cross-subnetwork visibility.

Furthermore, to demonstrate the functionality of the CRSF, we mathematically model a system of interconnected subnetworks and formulate an optimization framework for service function (SF) selection. The framework employs a \textit{priority-weighted, Quality of Service (QoS)-aware model} that selects the SF which offers the highest achievable QoS, while incorporating request-specific preferences towards particular subnetworks through priority weights. Our simulation results confirm the effectiveness of this balanced approach, showing that it consistently outperforms a baseline strategy that relies solely on priorities, thereby achieving a significantly higher aggregate QoS.

The remainder of this paper is organized as follows. Section~\ref{sec:systemDesign} presents the design scenario and the proposed 6G architecture with the CRSF, including a detailed call-flow description of its functionality. Section~\ref{sec:systemModel} describes the mathematical system model and the optimization formulation. Section~\ref{sec:evaluation} outlines the simulation setup, evaluation metrics, and baseline, followed by a detailed performance analysis. Section~\ref{sec:relatedWork} provides a review of related literature, and Section~\ref{sec:conclusion} concludes the paper.

\section{System Design}\label{sec:systemDesign}

In this section, we describe the scenario chosen to design the 6G system architecture and introduce our proposed CRSF with a detailed call flow describing its integration.

\begin{figure}[t!]
    \centering
    \includegraphics[width=\columnwidth, trim={0.1cm 4.5cm 0.3cm, 2cm},clip]{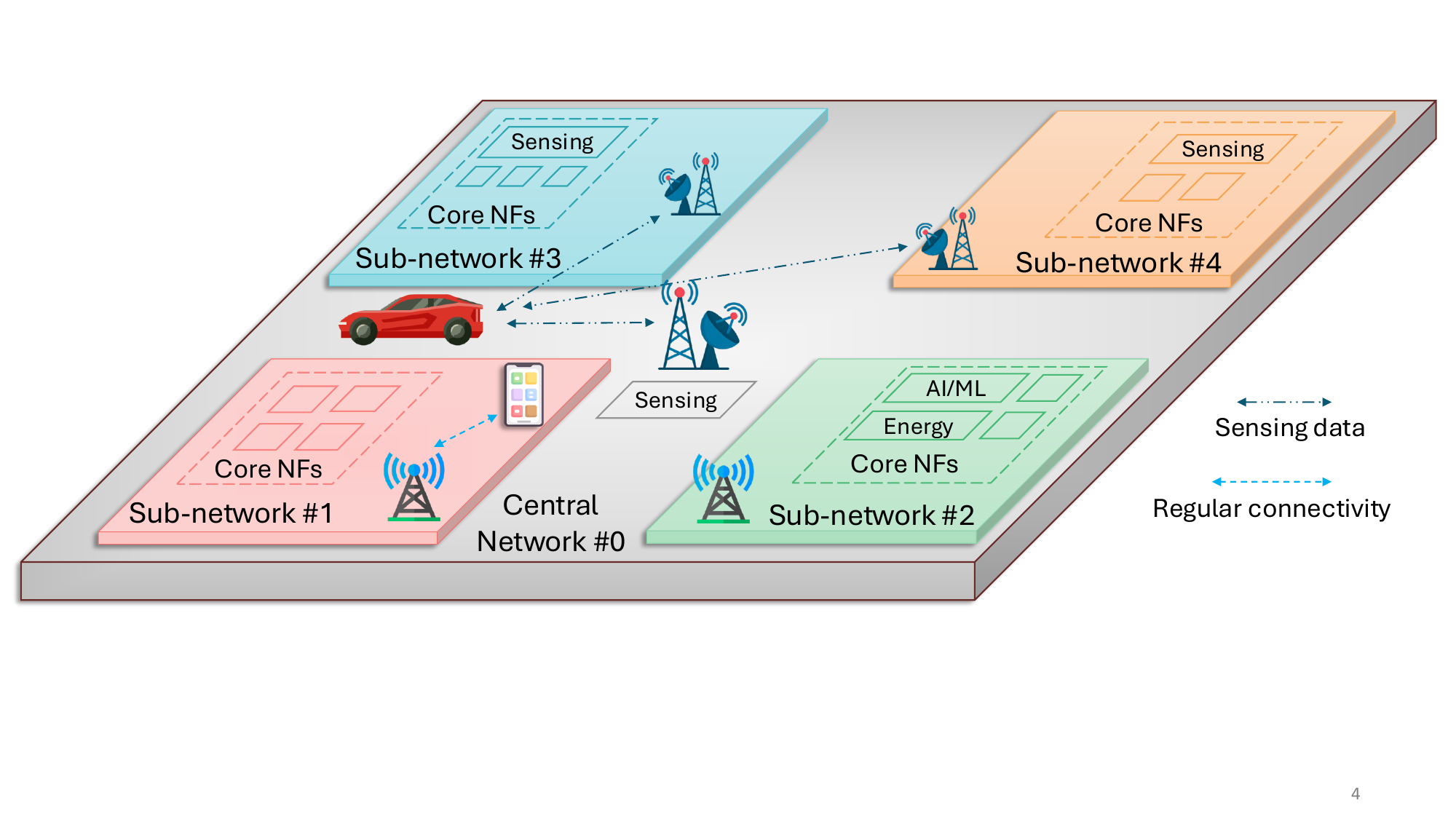}
    \caption{Design scenario with interconnected subnetworks alongside a central network offering beyond-connectivity services like sensing, AI/ML etc.}
    \label{fig:designScenario}
\end{figure}

\subsection{Scenario for System Design}
We illustrate the motivating use case behind our proposed 6G architecture in Figure~\ref{fig:designScenario}. In this scenario, multiple geographically isolated subnetworks are deployed alongside a central network. This setup is equivalent to a real-world enterprise deployment as shown in Figure~\ref{fig:vertical_usecase}, where different campuses of a company represent subnetworks, while the corporate headquarters acts as the central network. Each subnetwork is assumed to support standalone connectivity services for regular communication, but also participates in inter-subnetwork service sharing to enable beyond-connectivity functionalities such as sensing, energy control, AI/ML-based analytics, and localization. 

In this scenario, the User Equipment (UE) in subnetwork \#1 issues a request for sensing information about a target. However, the subnetwork \#1 does not natively provide sensing services. Thus, the request must be served by surrounding subnetworks that offer sensing capabilities--- specifically, subnetwork \#3, subnetwork \#4, and the central network \#0. To accomplish this, the UE must first discover which subnetworks provide the requested service (sensing in this case) and then obtain an intelligent selection of the most suitable subnetwork for request fulfillment. 

\subsection{Proposed Enhancement to 6G Service Architecture}
\begin{figure}[t!]
    \centering
    \includegraphics[width=\columnwidth, trim={3cm 3cm 3cm, 4cm},clip]{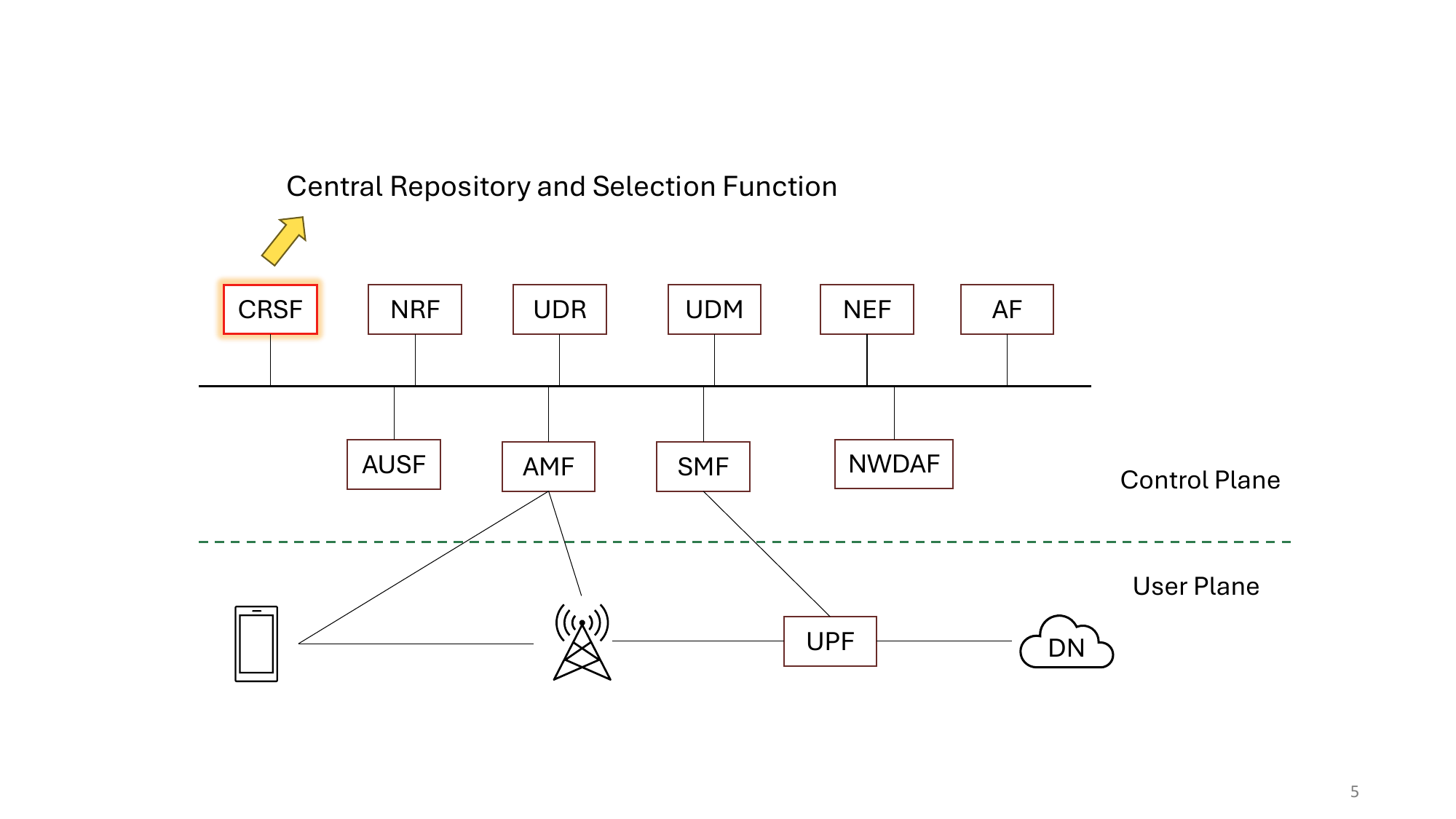}
    \caption{6G system architecture with CRSF alongside other network functions in the control plane of the central network}
    \label{fig:6garch}
\end{figure}

To enable this inter-subnetwork service discovery and selection, we propose a novel network function, called the \textbf{Central Repository and Selection Function (CRSF)}, to be deployed within the control plane of the central network alongside existing 5G/6G core network functions as shown in Figure~\ref{fig:6garch}. The role of the CRSF is twofold: (1) it maintains a global repository of available beyond-connectivity services offered by all associated subnetworks, and (2) it performs intelligent selection of the most suitable SF based on the incoming request's requirements and prevailing network conditions. 

We choose a centralized design for the CRSF for low complexity since a single logical entity can coordinate service discovery and selection without requiring distributed consensus protocols among subnetworks. In real deployments, the Public Land Mobile Network (PLMN) could serve as the central network that interconnects enterprise or factory subnetworks, making it a practical location to host the CRSF.

\begin{figure*}[t!]
    \centering
    \includegraphics[width=0.8\textwidth, trim={1.9cm 1.6cm 2.7cm, 1cm},clip]{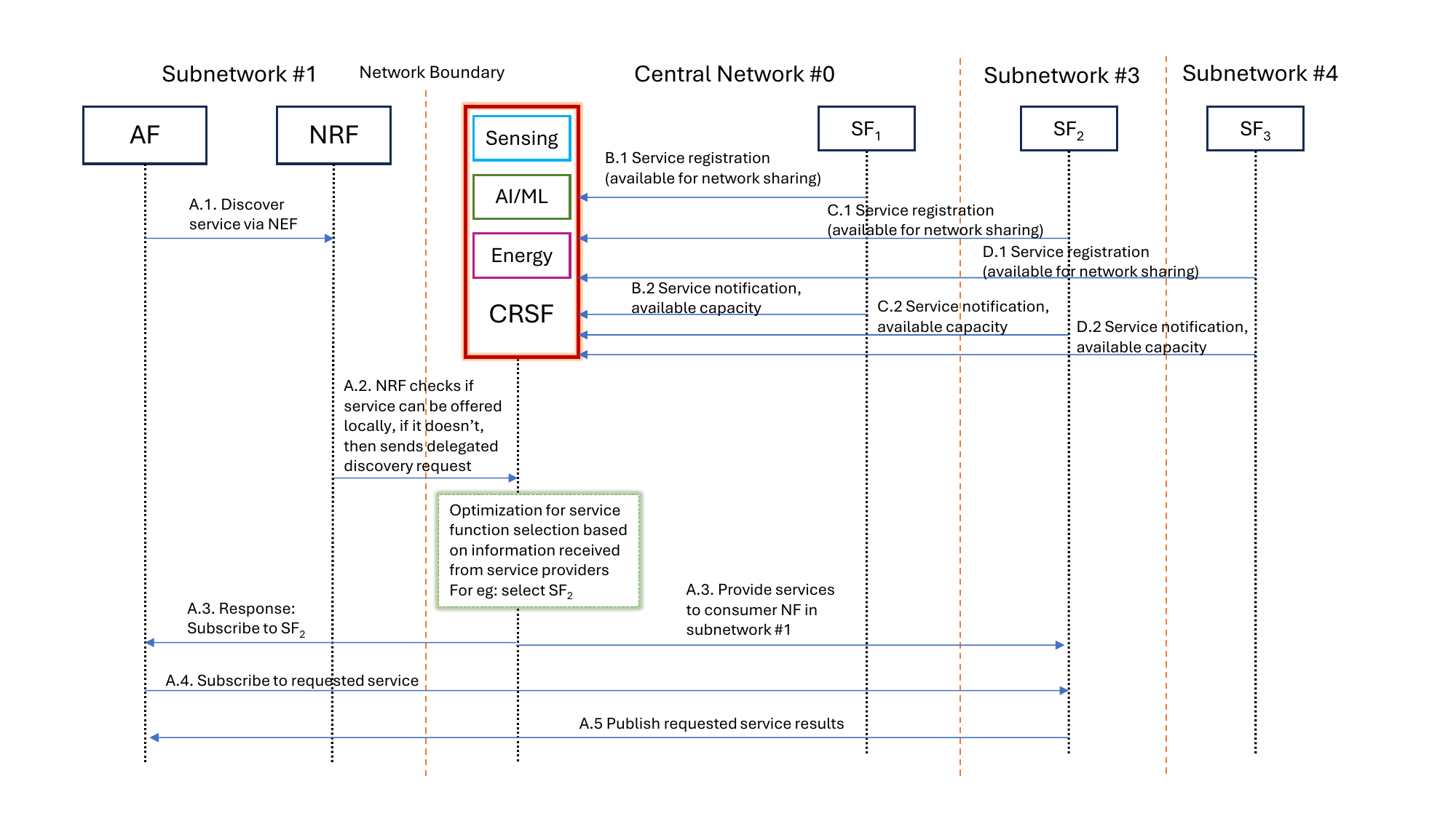}
    \caption{Call flow of the proposed 6G core network with the CRSF. The CRSF receives service registrations and capacity updates from subnetworks, performs optimization-based SF selection, and facilitates inter-subnetwork service discovery and selection.}
    \label{fig:callflow}
\end{figure*}

Figure~\ref{fig:callflow} illustrates a call flow sequence for the completion of the inter-network service discovery and selection procedure with the CRSF located in the central network, for the scenario depicted in Figure~\ref{fig:designScenario}. In a microservice based architecture, the CRSF could internally host different microservices, each specialized in acting as a service repository and performing selection for a particular type of beyond-connectivity service. The procedures involved in the functional integration of CRSF for a service connection within the 6G core are: 

\textit{Service Discovery}: \textbf{(A.1)} An Application Function (AF) (on behalf of the UE) in subnetwork \#1 initiates a service discovery request with the Network Repository Function (NRF) via the Network Exposure Function (NEF). \textbf{(A.2)} The NRF first checks whether the requested service can be fulfilled locally. If not, it delegates the discovery request to the CRSF in the central network.

\textit{Service Registration}: \textbf{(B.1, C.1, D.1)} Each SF ($\text{SF}_1, \text{SF}_2, \text{SF}_3$) registers itself with the CRSF, declaring availability for inter-network service sharing. \textbf{(B.2, C.2, D.2)} Along with registration, the SFs provide their service capabilities, such as QoS parameters and available load capacity to the CRSF. This aggregated service catalog enables the CRSF to maintain global visibility across subnetworks.

\textit{Service Selection}: Based on received service information (QoS, capacity), the CRSF selects the optimal SF to serve the incoming request (e.g. $\text{SF}_2$). \textbf{(A.3)} The CRSF communicates this selection decision to the service consumer, i.e., AF in subnetwork \#1 and the service provider, i.e., $\text{SF}_2$. 

\textit{Service Subscription}: \textbf{(A.4)} The AF subscribes to the chosen SF (e.g. $\text{SF}_2$) directly, and \textbf{(A.5)} the subscribed service results are published back to the requesting AF.

\section{Modeling and Problem Formulation}\label{sec:systemModel}

In this section, we mathematically model the system and formulate an optimization framework for the optimal selection of SFs at the CRSF.

\subsection{Service Request and Service Function Modeling}
We assume each available subnetwork hosts exactly one SF dedicated to any beyond-connectivity service. Thus, the number of SFs of a particular service type is equal to the number of available subnetworks offering that service. We denote the set of available SFs as $\mathcal{M}=\{m|m=1,\dots,|\mathcal{M}|\}$. The CRSF is assumed to operate in a time-slotted manner, where each time slot corresponds to the interval during which the CRSF performs the optimal selection of SFs for a batch of incoming requests. Let $\mathcal{R}=\{r|r=1,\dots,|\mathcal{R}|\}$ denote the set of requests arriving at the beginning of a time slot. Thus, the SF selection task at the CRSF is equivalent to an assignment problem where each request $r \in \mathcal{R}$ must be assigned to one of the available SFs in $\mathcal{M}$.

Since the available SFs reside within different subnetworks, potentially belonging to distinct administrative domains, each incoming request may assign different levels of preference to these domains. We represent these different preferences as priority-weights in our model. For a request $r \in \mathcal{R}$, the priority-weight $S_{r,m}$ reflects its preference towards subnetwork hosting SF $m \in \mathcal{M}$. For instance, if the request $r$ originates from a network within the same trust domain as the subnetwork of SF $m$, then $S_{r,m}$ could be a high value, indicating a higher priority for selecting the SF $m$ during the service selection/request assignment process.

Additionally, we assume that each incoming service request belongs to a specific service category. The number of service categories depends on the type of beyond-connectivity service. For example, a sensing service request may belong to categories such as object detection and tracking, environmental monitoring, or motion monitoring (as specified in Table 6.2-1 of 3GPP TS 22.137\cite{3gppTS22137}). Let the set of service categories be $\mathcal{K} = \{k | k = 1, \dots, |\mathcal{K}|\}$. We then introduce the binary indicator variable $\gamma_{r,k}$ defined as
\begin{equation}
    \gamma_{r,k} = \begin{cases}
    1, & \text{if request $r$ has service category $k$}.\\
    0, & \text{otherwise}.
  \end{cases}
  \label{eq:gamma}
\end{equation}

As mentioned in Section~\ref{sec:systemDesign}, the CRSF receives updated information about the QoS metrics and available load capacity of SFs. This information is refreshed at the beginning of each processing interval to ensure that decisions are based on the most recent network state. Let there be $N$ parameters that characterize the QoS metrics of a given service type. For example, in the case of a sensing service, these parameters may include position accuracy, probability of detection, false alarm rate, and others (see Table 6.2-1 in \cite{3gppTS22137}). 
The overall QoS provided by an SF can be expressed as a weighted sum of these parameters. Specifically, let $p_{n,m}$ denote the value of the $n^{th}$ parameter where $n \in \{1, \dots, N\}$ reported by SF $m$, and let $w_{k,n}$ be the weight assigned to the $n^{th}$ parameter for a request belonging to service category $k$. Then, the QoS score achieved at SF $m$  with respect to category $k$ is defined as
\begin{equation}
    Q_{k,m} = \sum_{n=1}^N w_{k,n} p_{n,m} 
    \label{eq:qos_km}
\end{equation}

Accordingly, the QoS score achieved by a specific request $r$ when served by SF $m$ can be expressed as
\begin{equation}
    \mathcal{Q}_{r,m} = \sum_{k \in \mathcal{K}}\gamma_{r,k}Q_{k,m}   
    \label{eq:qos_rm}
\end{equation}
where the indicator variable $\gamma_{r,k}$ ensures that only the QoS of the relevant service category contributes to the evaluation.

Finally, we introduce the binary decision variable $\alpha_{r,m}$, which captures the assignment of requests to SFs:
\begin{equation}
    \alpha_{r,m} = \begin{cases}
    1, & \text{if request $r$ is assigned to SF $m$}.\\
    0, & \text{otherwise}.
  \end{cases}
  \label{eq:alpha}
\end{equation}

\begin{table}[b!]
    \caption{Summary of Notations}
    \begin{minipage}{0.97\columnwidth}
    \centering
    \begin{tabular}{>{\raggedright\arraybackslash}p{0.13\columnwidth} p{0.73\columnwidth}}
    \toprule
    \textbf{Symbol} & \textbf{Description} \\
    \midrule
        $\alpha_{r,m}$ & Binary decision variable indicating if request $r$ is served by SF $m$ \\
        $\gamma_{r,k}$ & Binary variable (known) indicating if request $r$ belongs to category $k$\\
        $p_{n,m}$ & Value of $n^{th}$ QoS parameter at SF $m$\\
        $w_{k,n}$ & Weight assigned to $n^{th}$ QoS parameter for request of category $k$ \\
        $Q_{k,m}$ & Weighted aggregate of QoS for a request of category $k$ at SF $m$ \\
        $\mathcal{Q}_{r,m}$ & Weighted aggregate of QoS for request $r$ at SF $m$ \\
        $S_{r,m}$ & Weight reflecting priority of request $r$ towards administrative domain of SF $m$ \\
        $V^\text{Priority-QoS}_r$ & Priority-weighted QoS value for request $r$ \\
        $L_{r,m}$ & Service communication latency incurred by request $r$ at SF $m$ \\
        $T_k$ & Latency threshold for request of category $k$\\
        $U_k$ & Resource utilization for request of category $k$\\
        $C_m$ & Capacity of SF $m$ \\ 
        $\mathcal{R}$ & Set of incoming service requests \\ 
        $\mathcal{M}$ & Set of available service functions \\ 
        $\mathcal{K}$ & Set of service categories \\
        $N$ & Number of service parameters reflecting QoS metrics\\ 
           
    \bottomrule
    \end{tabular}
    \end{minipage}
    \label{tab:symbols}
\end{table}

\subsection{Optimization Problem Formulation}

The objective of the CRSF is to assign each incoming request to a suitable SF such that high-priority SFs are selected while maximizing the QoS achieved.
For a request $r$, we define a joint priority-weighted QoS value as:
\begin{equation}
    V^\text{Priority-QoS}_r = \sum_{m \in \mathcal{M}}\alpha_{r,m}S_{r,m}\mathcal{Q}_{r,m}
    \label{eq:obj}
\end{equation}

The overall objective is to maximize the total priority-weighted QoS $\forall r \in \mathcal{R}$.
\begin{equation}
    \textbf{P1:} \quad \max_{\alpha_{r,m}} \sum_{r \in \mathcal{R}} \sum_{m \in \mathcal{M}}\alpha_{r,m}S_{r,m}\mathcal{Q}_{r,m} \notag 
\end{equation}

\subsubsection{Latency Constraint}
Let $L_{r,m}$ denote the service communication latency incurred if a request $r$ is served by SF $m$, and let $T_k$ be the latency threshold for service category $k$. The service communication latency $L_{r,m}$ incurred by any request $r$ belonging to category $k$, when served by SF $m$, should not exceed its threshold $T_k$.
\begin{equation} \label{eq:constraint1}
    \sum_{m \in \mathcal{M}} \alpha_{r,m}L_{r,m} \leq \sum_{k \in \mathcal{K}} \gamma_{r,k}T_{k} \quad \forall r \in \mathcal{R} \tag{C1} \\
\end{equation}

\subsubsection{Capacity Constraint}
Each SF $m$ has a load capacity $C_m$, which is notified to the CRSF at the start of each time slot. Let $U_k$ be the resource utilization required by a request of category $k$. The combined resource utilization of all requests assigned to SF $m$ must not exceed its capacity $C_m$. 
\begin{equation}\label{eq:constraint2}
    \sum_{r \in \mathcal{R}} \sum_{k \in \mathcal{K}} \alpha_{r,m}\gamma_{r,k}U_{k} \leq C_m \quad \forall m \in \mathcal{M} \tag{C2} \\ 
\end{equation}

\subsubsection{Assignment Constraint}
Each request can be assigned to at most one SF. 
\begin{equation} \label{eq:constraint3}
    \sum_{m \in \mathcal{M}} \alpha_{r,m} \leq 1 \quad \forall r \in \mathcal{R}  \tag{C3}
\end{equation}

In summary, the optimization problem is a binary linear program (BLP) with binary decision variables $\alpha_{r,m}$, for all requests $r\in \mathcal{R}$ and SFs $m\in \mathcal{M}$:  
\begin{align}
    \max_{\alpha_{r,m}} \quad & \sum_{r \in \mathcal{R}} \sum_{m \in \mathcal{M}}\alpha_{r,m}S_{r,m}\mathcal{Q}_{r,m} \tag{P1} \label{eq:optProblem}\\
\text{subject to} \quad  & \eqref{eq:constraint1}, \eqref{eq:constraint2}, \eqref{eq:constraint3} \notag
\end{align}

\section{Evaluations}\label{sec:evaluation}

In this section, we present the evaluation of our modeled framework using simulations. We begin with the description of the simulation setup tailored to the sensing service selection scenario. Next, we define the performance metrics used to analyze system behavior and describe the baseline scheme used for comparison. Finally, we analyze the performance of our proposed approach against the baseline.

\subsection{Simulation Setup}

While our system model described in Section~\ref{sec:systemModel} is applicable to any beyond-connectivity service type, we design our experiments specifically for the sensing service.
We choose 6 parameters ($N=6$), listed in Table~\ref{tab:sensingParams}, to characterize the sensing QoS at each SF. Additionally, we assume 5 distinct categories ($\mathcal{|K|}=5$) of the sensing service type.
These parameters and categories are chosen in accordance with the 3GPP specifications in TS 22.137~\cite{3gppTS22137}. 
The variables used in Equations \eqref{eq:obj}, \eqref{eq:constraint1}, \eqref{eq:constraint2} are sampled from the ranges specified in Table~\ref{tab:simParams}. We conduct Monte-Carlo simulations in Python by sampling the input variables from uniform distributions within the ranges given in Tables~\ref{tab:sensingParams} and~\ref{tab:simParams} and vary the number of incoming requests and available SFs to study the resulting system performance. 
The optimization problem \eqref{eq:optProblem} is NP-Hard in complexity which we solve using the Gurobi optimization solver~\cite{gurobi} to generate an optimal solution in each simulation round. 100 simulation rounds are conducted to obtain each data point.

\begin{table}[b!]
    \centering
    \caption{Sensing QoS parameters to calculate $Q_{k,m}$}
    \begin{tabular}{lr}
        \toprule
        QoS Parameter (unit) & [Min, Max] \\
        \midrule
        Position accuracy (cm) & $[20, 500)$ \\
        Latency (ms) & $[10, 500)$ \\
        Sensing Range (m) & $[50, 300)$ \\
        Resolution (cm) & $[1, 20)$ \\
        Detection Probability & $(0.5, 1)$ \\
        False Alarm Probability & $(0, 0.1)$ \\
        \bottomrule
    \end{tabular}
    \label{tab:sensingParams}
\end{table}

\begin{table}[b!]
    \centering
    \caption{Simulation Values for Variables} 
    \begin{tabular}{lr}
        \toprule
        Variable (Expression) &  [Min, Max]  \\
        \midrule
        QoS parameter weight $(w_{k,n})$ & $(0,1]$ \\
        Priority weight $(S_{r,m})$ & $[1, 10]$ \\
        Latency $(L_{r,m})$ & $[60, 120)$ \\
        Latency Threshold $(T_k)$  & $[90, 140)$ \\
        Capacity $(C_m)$ & $[30, 50]$ \\
        Utilization $(U_k)$ & $[5, 10]$ \\
        \bottomrule
    \end{tabular}
    \label{tab:simParams}
\end{table}


\subsection{Evaluation Metrics and Baseline}
To evaluate the effectiveness of our proposed framework, we consider three performance metrics.
\begin{itemize}
    \item \textbf{Aggregate priority-weighted QoS}: $\sum_{r \in \mathcal{R}} V_r^{\text{Priority-QoS}}$, representing the total priority-weighted QoS achieved across all requests.
    \item \textbf{Assignment Success Rate (ASR)}: the ratio of successfully assigned requests to the total number of requests.
    \item \textbf{Average priority-weighted QoS per request}: defined as $\frac{\sum_{r\in \mathcal{R}} V_r^{\text{Priority-QoS}}}{|\mathcal{R}|}$ and evaluated when ASR equals 100\%.
\end{itemize}

For performance comparison, we define a baseline strategy for request assignment. In the baseline, requests are allocated exclusively according to their priorities, with each request mapped to the highest-priority SF. Since our proposed solution accounts for latency, capacity, and assignment constraints, we impose the same on the baseline to ensure fair comparison. Under this setting, the baseline objective reduces to:
\begin{equation}
    \max_{\alpha_{r,m}} \sum_{r\in \mathcal{R}} \sum_{m\in \mathcal{M}} S_{r,m} \notag
\end{equation}

subject to the constraints \eqref{eq:constraint1}, \eqref{eq:constraint2}, \eqref{eq:constraint3}.

The subsequent plots report these three metrics for both the proposed approach and the baseline.

\begin{figure}[tb!]
    \centering
    \begin{subfigure}[h!]{0.49\columnwidth}
        \centering
        \includegraphics[width=\linewidth]{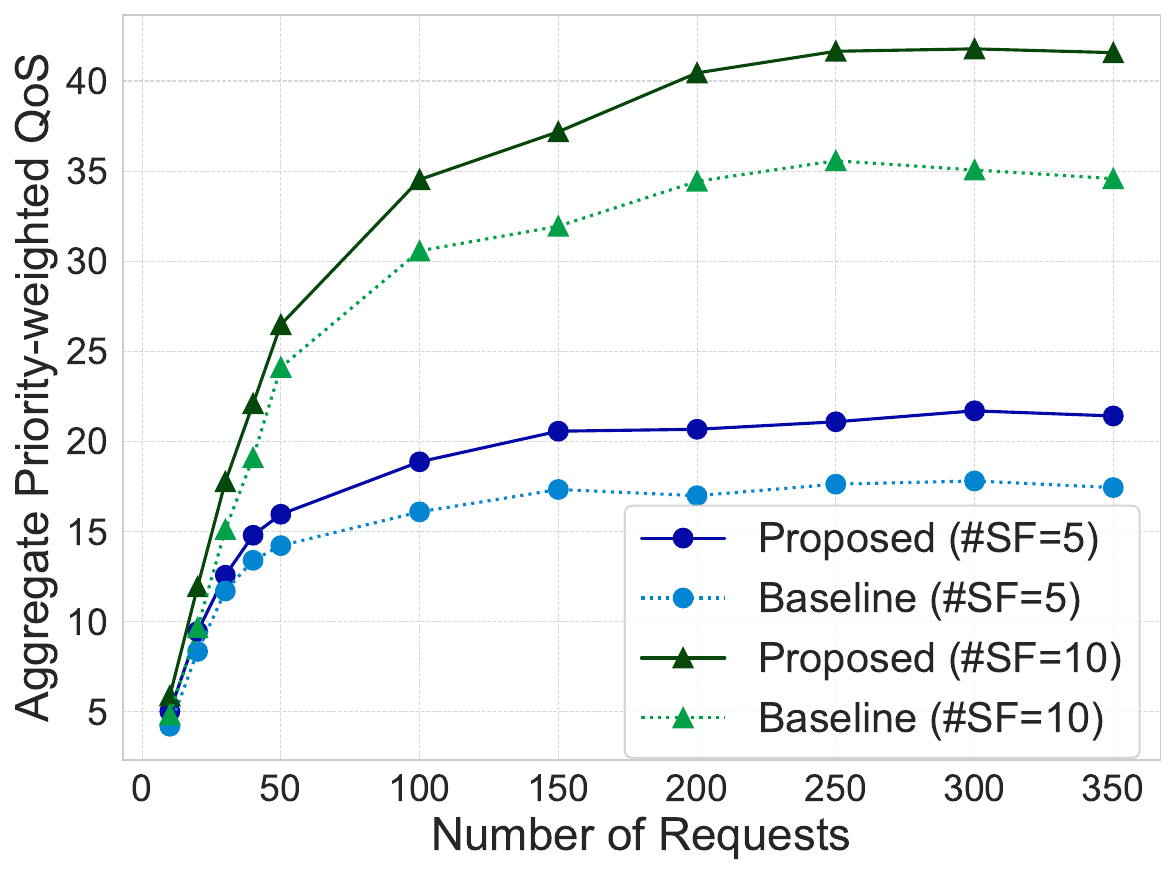}
        \vspace{-6mm}
        \caption{}
    \end{subfigure}
    \centering
    \begin{subfigure}[h!]{0.49\columnwidth}
        \centering
        \includegraphics[width=\linewidth]{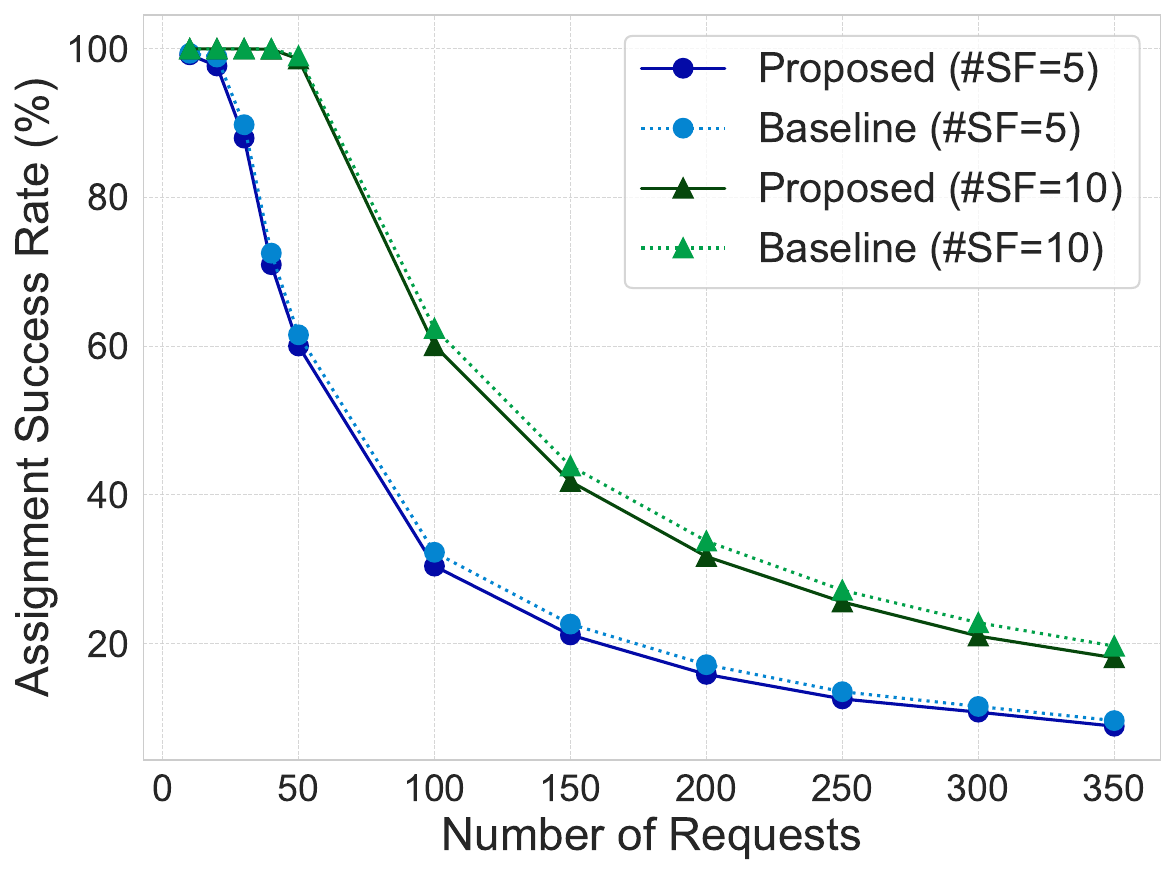}
            \vspace{-6mm}
        \caption{}
    \end{subfigure}
    \caption{Performance comparison of the proposed optimization framework and the baseline with increasing number of requests, for number of SFs = \{5,10\}}
    \label{fig:result-1}
\end{figure}

\begin{figure}[tb!]
    \centering
    \begin{subfigure}[h!]{0.49\columnwidth}
        \centering
        \includegraphics[width=\linewidth]{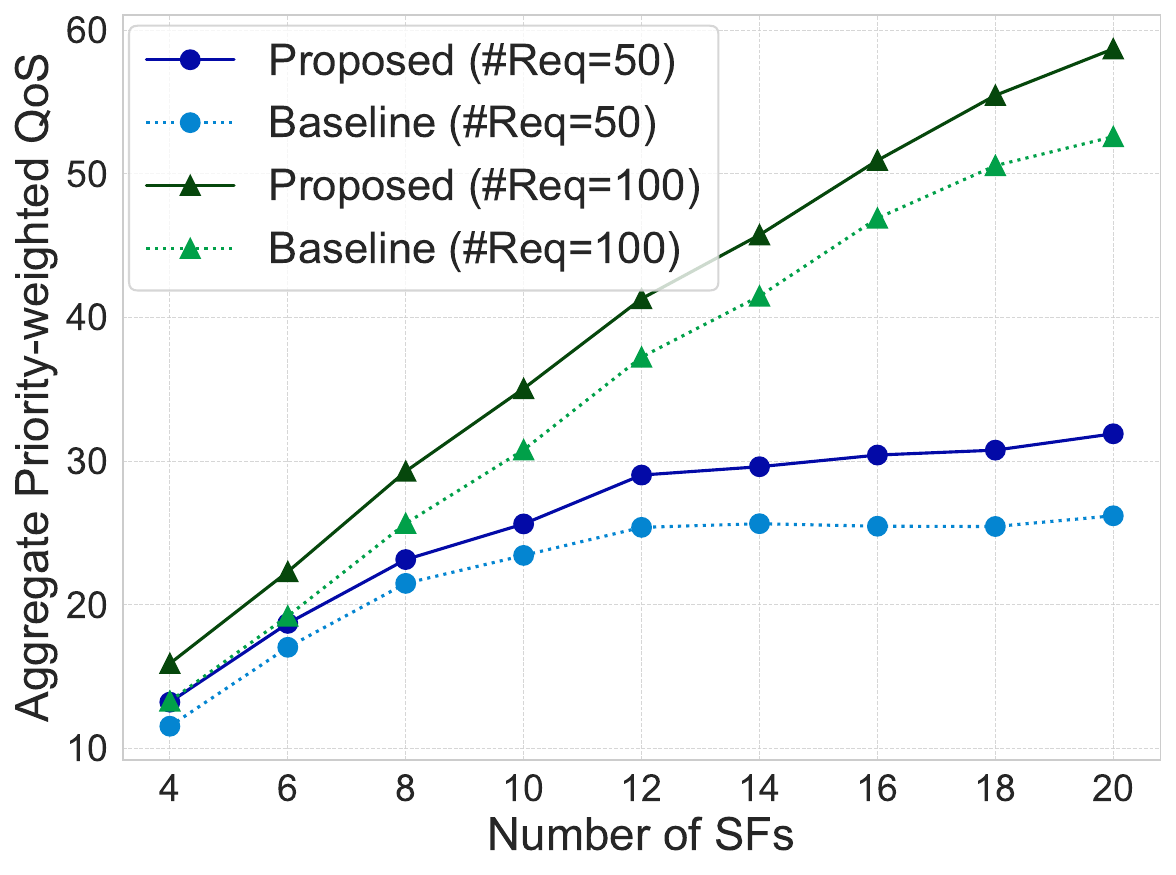}
        \vspace{-6mm}
        \caption{}
    \end{subfigure}
    \centering
    \begin{subfigure}[h!]{0.49\columnwidth}
        \centering
        \includegraphics[width=\linewidth]{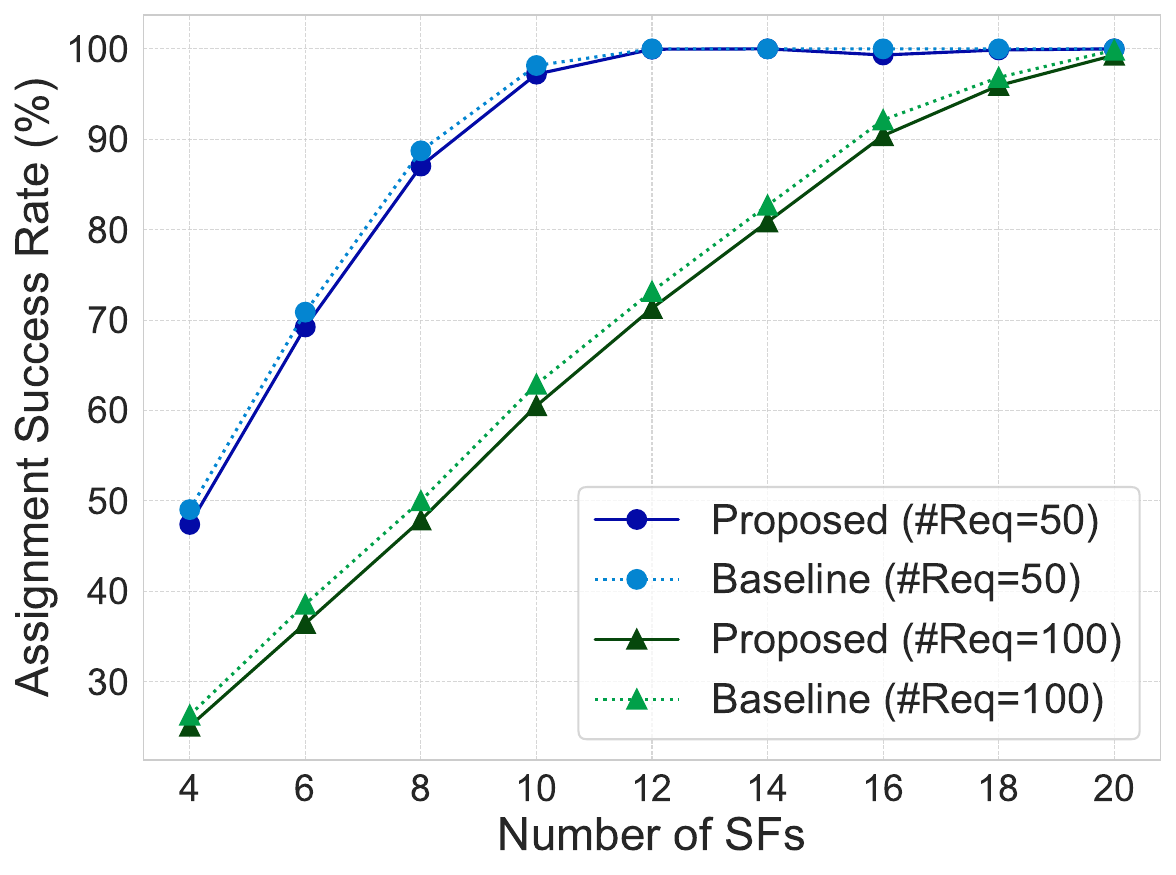}
        \vspace{-6mm}
        \caption{}
    \end{subfigure}
    \caption{Performance comparison of the proposed optimization framework and the baseline with increasing number of SFs, for number of requests = \{50,100\}}
    \label{fig:result-2}
    \vspace{-4mm}
\end{figure}

\subsection{Performance Analysis}
In Figure~\ref{fig:result-1}, we compare the proposed optimization scheme with the baseline approach in terms of ASR and aggregate priority-weighted QoS with an increasing number of requests. Figure~\ref{fig:result-1}(a) shows that the proposed scheme consistently yields higher aggregate priority-weighted QoS compared to the baseline, for both number of SFs = \{5, 10\}, with the benefit becoming more pronounced as the number of requests increases. As shown in Figure~\ref{fig:result-1}(b), the ASR decreases with increasing request load due to the limited number of SFs. The aggregate priority-weighted QoS ultimately saturates due to load saturation in a resource-constrained system.
The baseline achieves marginally higher ASR since it assigns requests solely based on priority, leading to higher numbers of feasible assignments under the strict latency and capacity constraints. However, the difference in ASR is negligible in contrast to the gains achieved in overall QoS in our proposed approach.

Figure~\ref{fig:result-2} shows the trend of aggregate priority-weighted QoS and ASR with increasing number of available SFs for number of requests = \{50, 100\}. Figure~\ref{fig:result-2}(a) demonstrates a higher QoS in the proposed scheme compared to the baseline, with increasing gain consistent across different number of requests. As shown in Figure~\ref{fig:result-2}(b), ASR improves with the number of SFs since additional resource availability increases the likelihood of successful request allocation, with ASR becoming 100\% for a large number of SFs.
Both Figure~\ref{fig:result-1} and Figure~\ref{fig:result-2} also aid in identifying the scalability limits of the system under a chosen set of design variables, which is a critical consideration for resource-constrained networks.

\begin{figure}[tb!]
    \centering
    \includegraphics[width=0.55\columnwidth]{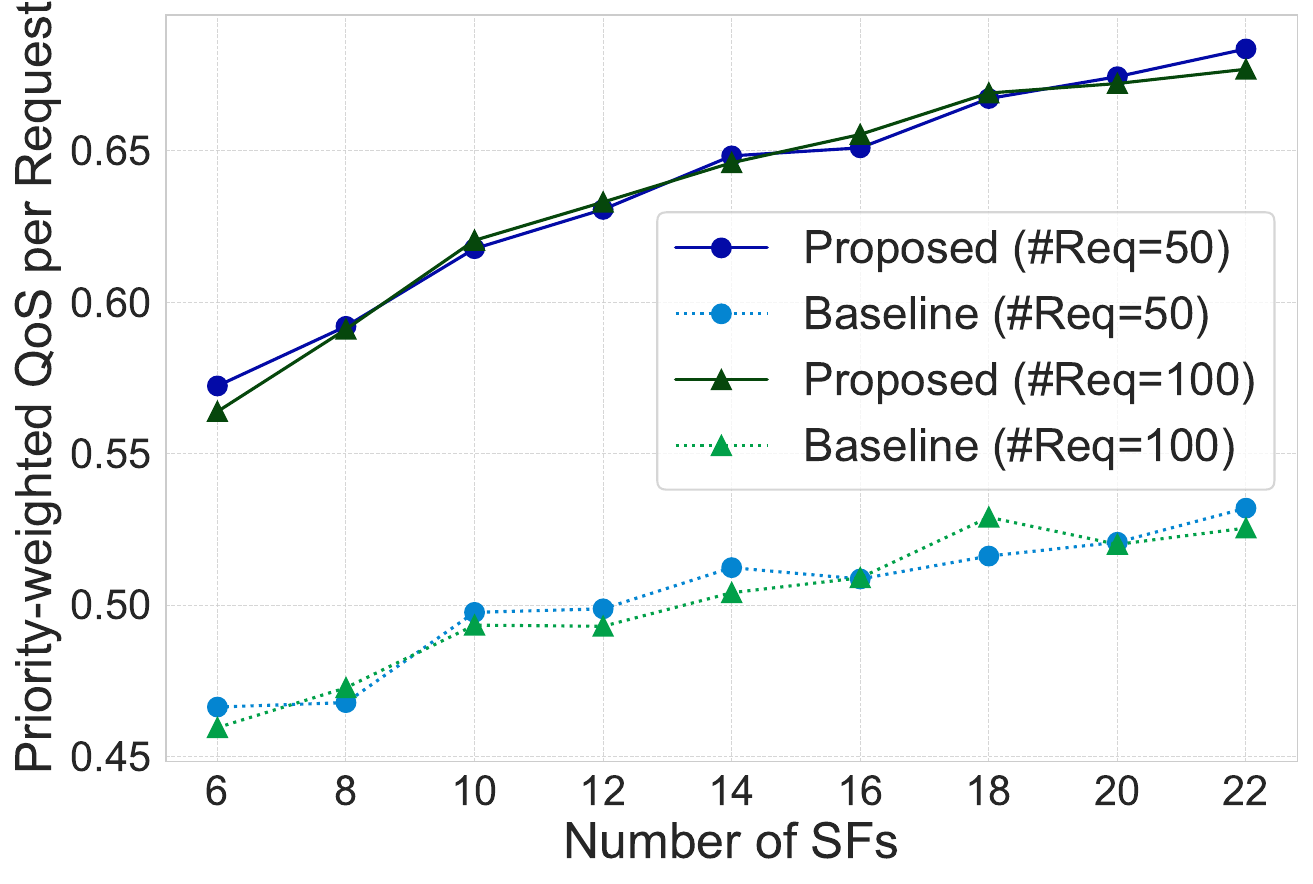}
    \caption{Priority-weighted QoS per request when assignment success rate is 100\%}
    \label{fig:result-3}
\end{figure}

Additionally, we observe how the priority-weighted QoS per request evolves as the number of SFs increases, under the condition of ASR=100\% in Figure~\ref{fig:result-3}. When all the requests are successfully assigned (ASR=100\%), a larger pool of SFs expands the space of feasible request-to-SF assignments. This, in turn, enables the selection of better matches and results in higher priority-weighted QoS per request.

\begin{figure}[tb!]

    \begin{subfigure}[h!]{0.49\columnwidth}
        \centering
        \includegraphics[width=\linewidth]{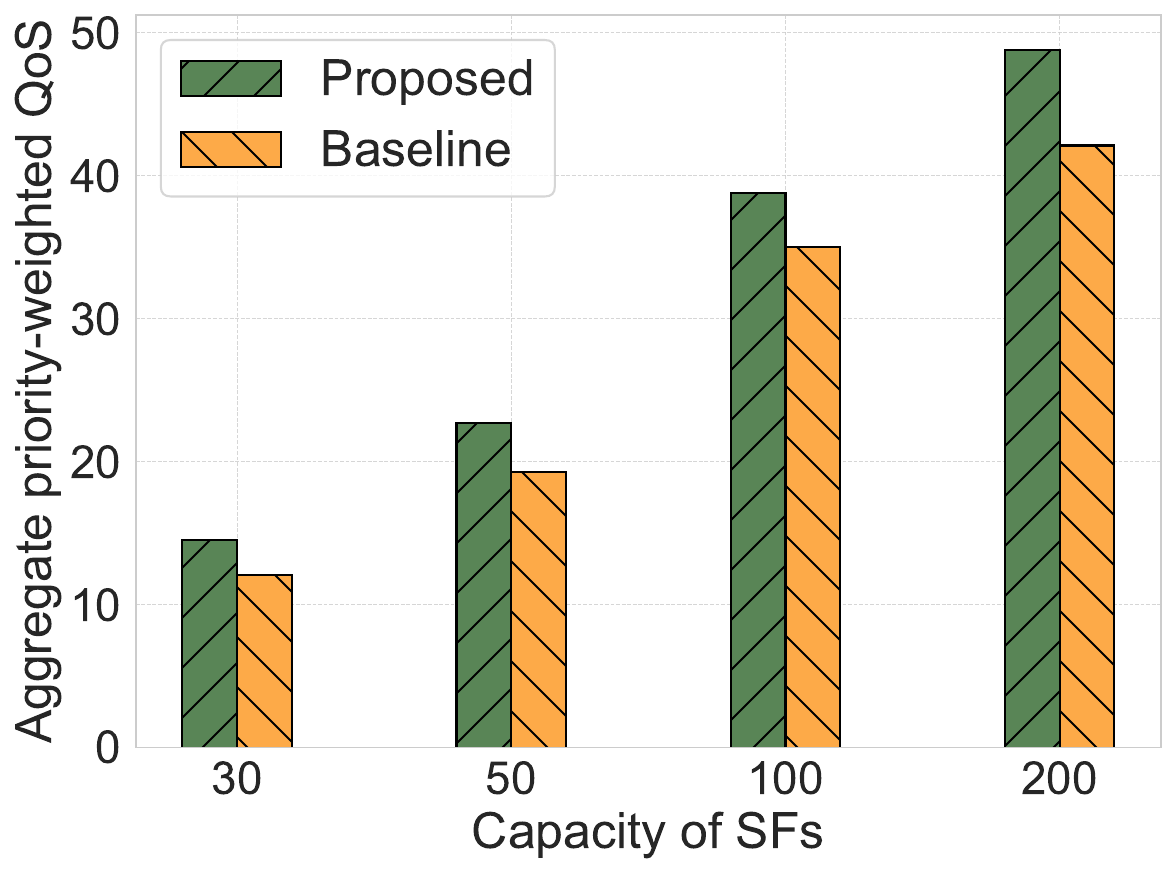}
            \vspace{-6mm}
        \caption{}
    \end{subfigure}
    \begin{subfigure}[h!]{0.5\columnwidth}
        \centering
        \includegraphics[width=\linewidth]{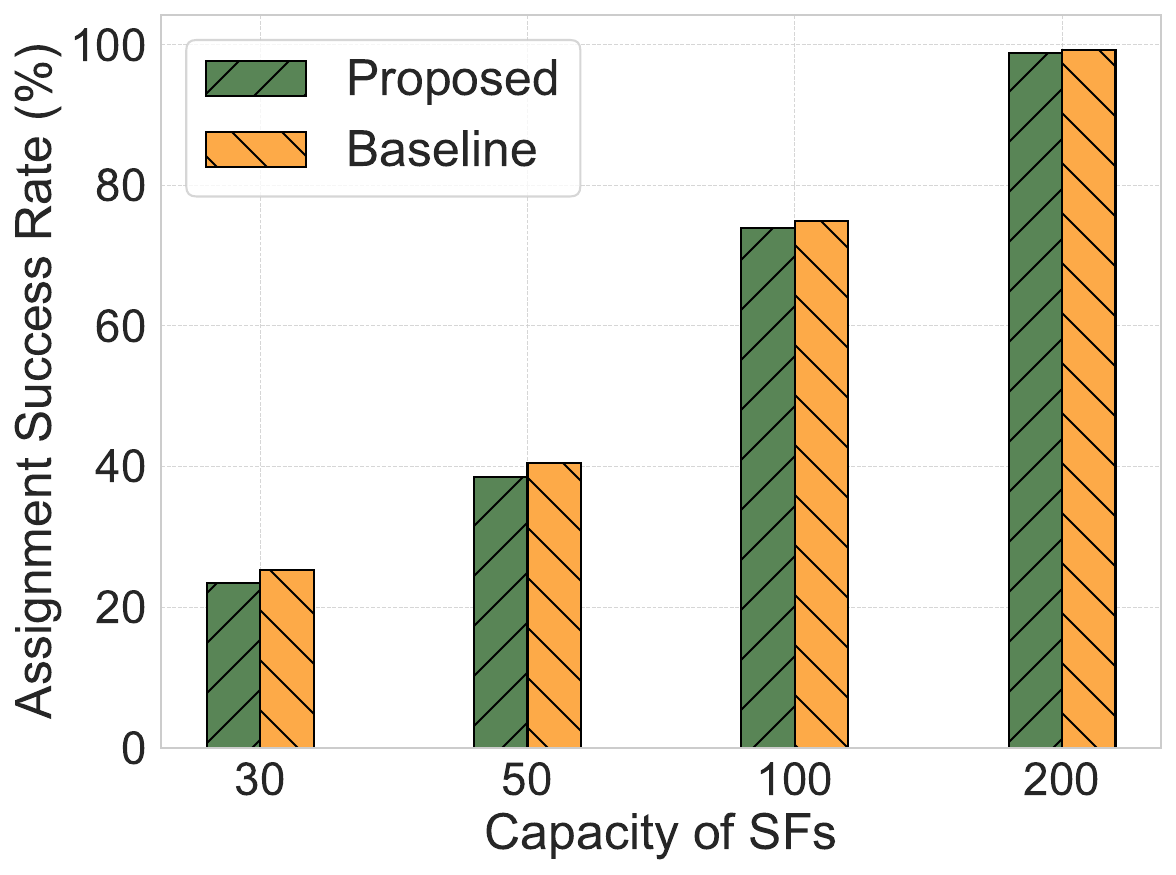}
            \vspace{-6mm}
        \caption{}
    \end{subfigure}
    \caption{Effect of Capacity of SFs on aggregate priority-weighted QoS and ASR when number of SFs = 5 and number of requests = 100}
    \label{fig:result-4}
    \vspace{-4mm}
\end{figure}


We show the effect of changing SF capacity (as modeled in the constraint~\ref{eq:constraint2}) on aggregate priority-weighted QoS and ASR in Figure~\ref{fig:result-4}. As seen in Figure~\ref{fig:result-4}(a), aggregate QoS improves consistently with increasing SF capacity, since larger capacities allow the system to accommodate more requests, leading to higher aggregate QoS. The proposed scheme achieves higher QoS across all capacity levels, and the performance gap with the baseline widens as capacity increases, indicating that the proposed method leverages additional resources more effectively. Figure~\ref{fig:result-4}(b) shows that ASR also rises with capacity and approaches 100\% once the system becomes sufficiently provisioned. These results highlight the advantage of the proposed scheme in delivering higher service quality, particularly in resource-rich environments.

\section{Related Literature}\label{sec:relatedWork}

While no existing work discusses service discovery and selection for shared services in 6G interconnected subnetworks, we present some interesting tangential works that provide unique solutions for service discovery and overall service-oriented architecture in 6G.

\textbf{Service Discovery Methods:}
\cite{SaRD} proposes a service and resource discovery architecture for the 6G edge-cloud continuum, integrating dynamic zoning, semantic networking, ML-based prediction, and blockchain for secure discovery.
\cite{bowei} presents a distributed service registration and discovery framework for 6G satellite networks using consistent hashing to reduce latency from concurrent registrations.
\cite{blockchainSD} introduces a blockchain-based distributed service registry for 6G core networks, leveraging hyperledger fabric to avoid single-point failures at the NRF in multi-operator private networks.

\textbf{6G service architecture:}
\cite{6gSBA_1} outlines a conceptual service-oriented architecture that unifies control plane functions of the core and radio access network within a single set of NFs, enabling network-of-networks connectivity and dynamic local subnetwork instantiation.
\cite{6gsba_2} presents a cognitive service architecture where edge and cloud cores collaborate through knowledge graphs for intelligent, adaptive service management.
\cite{6gsba_3} proposes decoupling tightly coupled sequential NF registration procedures into parallel transactions to mitigate signaling bottlenecks in 5G SBA.

\section{Conclusion}\label{sec:conclusion}

We introduced the CRSF, a novel core network function that facilitates the discovery and selection of beyond-connectivity services across interconnected subnetworks. We formulated this selection as a priority-weighted, QoS-aware optimization problem and demonstrated that our proposed method delivers superior performance in terms of aggregate QoS as compared to the chosen baseline. By enabling intelligent service selection, the CRSF lays the essential groundwork for standardized, service-centric 6G architectures toward collaborative and integrated networks. Future work will focus on integrating learning-based algorithms into the CRSF for more dynamic decision-making and exploring joint-objective optimizations to enhance its robustness for real-world deployments.


\bibliographystyle{IEEEtran}
\bibliography{references.bib}

\end{document}